\documentclass[twocolumn,showpacs,preprintnumbers,amsmath,amssymb]{revtex4}

\usepackage{graphicx}
\usepackage{dcolumn}
\usepackage{bm}
\usepackage{epsfig}

\begin{document}

\title{The 1D interacting anyon gas: low-energy properties and Haldane exclusion statistics}
\author{M.T. Batchelor$^1$,  X.-W. Guan$^1$ and N. Oelkers$^2$}
\affiliation{$^1$ Department of Theoretical Physics, Research School of Physical Sciences and Engineering, and\\
Mathematical Sciences Institute,
Australian National University, Canberra ACT 0200,  Australia\\
$^2$ Department of Mathematics, University of Queensland, Brisbane QLD 4072, Australia}

\date{\today}

\begin{abstract}
\noindent
The low energy properties of the one-dimensional anyon gas with $\delta$-function
interaction are discussed in the context of its Bethe ansatz solution. 
It is found that the anyonic statistical parameter and the dynamical coupling constant 
induce Haldane exclusion statistics interpolating between bosons and fermions. 
Moreover, the anyonic parameter may trigger statistics beyond Fermi statistics 
for which the exclusion parameter $\alpha$ is greater than one. 
The Tonks-Girardeau and the weak coupling limits are discussed in detail.
The results support the universal role of $\alpha$ in the dispersion relations.
\end{abstract}



\maketitle

Anyons, 
which are used to describe particles with generalized fractional statistics \cite{F-S,Haldane},
are becoming of increasing importance in condensed matter physics \cite{cond1} 
and quantum computation \cite{comp}. 
The concept of anyons provides a successful theory of the fractional quantum
Hall (FQH) effect \cite{QFHE}. 
In particular, the signature of fractional statistics has recently been observed in experiments
on the elementary excitations of a two-dimensional electron gas in the
FQH regime \cite{cond1}.  
These developments are seen as promising opportunities for further insight into the 
FQH effect, quantum computation, superconductivity and  other fundamental problems in quantum physics.

In one-dimension ($1$D), collision is the only way to interchange two particles. 
Accordingly, interaction and statistics are inextricably related in $1$D systems.
The $1$D Calogero-Sutherland model is seen to obey fractional
exclusion statistics \cite{Ha,Ha2}.  
In the sense of Haldane exclusion statistics, the $1$D interacting
Bose gas is equivalent to the ideal gas with generalized fractional statistics \cite{Wu,Isakov}.
We consider an integrable model of anyons with $\delta$-function interaction
introduced and solved by Kundu \cite{Kundu}.
Here we obtain  the low energy properties and Haldane
exclusion statistics of this 1D anyon gas.
We find that the low energies, dispersion relations and the generalized exclusion
statistics depend on both the anyonic statistical and the dynamical interaction parameters. 
The anyonic parameter not only interpolates between Bose and
Fermi statistics but can trigger statistics beyond Fermi statistics in
a super Tonks-Girardeau (TG)  gas-like phase.

{\em Bethe ansatz solution.}
We consider $N$ anyons with $\delta$-function
interaction in 1D with Hamiltonian \cite{Kundu}
\begin{eqnarray}
H &=&\frac{\hbar^2}{2m}\int_0^Ldx\partial \Psi ^{\dagger}(x) \partial
  \Psi(x)\nonumber\\
& &+\frac12 \, g_{\rm 1D} \int_0^Ldx\Psi^{\dagger}(x)\Psi^{\dagger}(x)\Psi(x) \Psi(x)
\label{Ham1}
\end{eqnarray}
and periodic boundary conditions (BC). 
Here $m$ denotes the atomic mass, 
$g_{\rm 1D}$ is the coupling constant and $x$ is a coordinate in length $L$. 
$\Psi ^{\dagger}(x)$ and $\Psi(x) $ are the creation and annihilation operators at point $x$  
satisfying the anyonic commutation relations
\begin{eqnarray}
\Psi (x_1)\Psi ^{\dagger}(x_2)&=&{\mathrm e}^{-\mathrm{i}\kappa
  w(x_1,x_2)}\Psi ^{\dagger}(x_2)\Psi(x_1)+\delta(x_1-x_2)\nonumber\\
\Psi ^{\dagger}(x_1)\Psi ^{\dagger}(x_2)&=&{\mathrm e}^{\mathrm{i}\kappa
  w(x_1,x_2)}\Psi ^{\dagger}(x_2)\Psi
^{\dagger}(x_1).
\end{eqnarray}
Here the multi-step function $w(x_1,x_2)=-w(x_2,x_1)=1$ for $x_1>x_2$, with $w(x,x)=0$.  
The coupling constant is determined by  $g_{\rm 1D}={\hbar^2c}/{m}$
where the coupling strength $c$ is tuned through an
effective $1$D scattering length $a_{1D}$ via  confinement in experiments.
Hereafter we set $\hbar =2m=1$ for convenience. 
We also use a dimensionless coupling constant $\gamma =c/n$ to
characterize different physical regimes of the anyon gas, where
$n=N/L$ is the linear density.

In contrast to the $1$D Bose gas
\cite{LL}, Hamiltonian (\ref{Ham1}) exhibits both anyonic statistical
and dynamical interactions, which can map into a $1$D interacting Bose
gas with multi-$\delta$-function and momentum-dependent interactions
\cite{Kundu}. In Ref.~\cite{Zoller} the authors have proposed a way to
observe the fractional statistics of anyons in a system of ultracold
bosonic atoms in a rapidly rotating trap.

Define a Fock vacuum
state $\Psi(x) \! \mid \! 0\rangle=0$ and assign all
particle coordinates $x_i$ in an order  
$x_1\leq x_2 \leq \cdots \leq x_N$. 
The $N$-particle eigenstate is 
\begin{equation}
\mid \! \Phi \rangle=\int_0^L dx^N {\mathrm e}^{-\mathrm{i}\frac{\kappa N}{2}}\chi
(x_1\ldots x_N)\Psi ^{\dagger}(x_1)\ldots \Psi
^{\dagger}(x_N) \! \mid \! 0\rangle
\label{state}
\end{equation}
where  the Bethe ansatz wave function is written as  
\begin{eqnarray}
\chi(x_1\ldots
x_N)&=&{\mathrm e}^{-\frac{\mathrm{i}\kappa}{2}\sum_{x_i<x_j}^Nw(x_i,x_j)}\sum_PA(k_{P1}\cdots
k_{PN})\nonumber\\
& & \times \, {\mathrm e}^{\mathrm{i}(k_{P1}x_1+\cdots +k_{PN}x_N)}.
\label{wave}
\end{eqnarray}
Here the sum extends over all $N!$ permutations $P$. In the
$N$-particle eigenstate the order with which the particles are created
incurs the phase factor in the wave function (\ref{wave}). Integration
involves the changes of the order in creating particles due to
the permutation of coordinates.  We easily see
that the wave function satisfies the anyonic symmetry $\chi(\cdots
x_i\cdots x_j \cdots )={\mathrm e}^{-\mathrm{i}\theta } \chi(\cdots
x_j\cdots x_i \cdots )$, in which the anyonic phase $\theta = \kappa
\left(\sum_{k=i+1}^jw(x_i,x_k)-\sum_{k=i+1}^{j-1}w(x_j,x_k) \right)$
for $i<j$.
We extract a global phase factor ${\mathrm e}^{-\mathrm{i}{\kappa N}/{2}}$ 
in order to symmetrize the anyonic phase factor in the wave
function (\ref{wave}) so that it has $\kappa \to \kappa +4\pi $
symmetry. The eigenstate still has $\kappa \to \kappa +2\pi $ symmetry.
However, the phase factors in the multi-valued wave function
(\ref{wave}) are diminished by those from permutations of the
particles in the eigenstate $\mid \!\! \Phi \rangle$ such that the
integrand in (\ref{state}) is single valued.

Solving the eigenvalue problem for Hamiltonian
(\ref{Ham1}) reduces to solving the quantum mechanics problem 
$H_N\chi (x_1\ldots x_N)=E\chi (x_1\ldots x_N)$, where  
\begin{equation}
H_N=-\frac{\hbar ^2}{2m}\sum_{i = 1}^{N}\frac{\partial
^2}{\partial x_i^2}+\,g_{\rm 1D}\sum_{1\leq i<j\leq N} \delta
(x_i-x_j)\label{Ham2}
\end{equation}
describes the $1$D $\delta$-function interacting quantum gas of $N$ anyons 
confined in a periodic length $L$.

\begin{center}
\begin{figure}[t]
\includegraphics[width=0.65\linewidth]{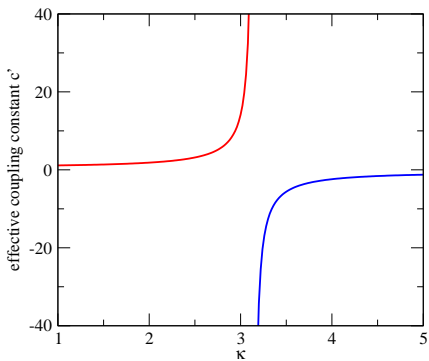}
\caption{The effective coupling constant $c'$ (in units of $c$) vs the anyonic parameter $\kappa$. 
A key feature of the model is that the anyonic statistical interaction induces a 
resonance-like behaviour where the interaction strength becomes very large.
}
\label{fig:c}
\end{figure}
\end{center}

The $N!$ coefficients $A(k_{P1} \ldots k_{PN})$ are obtained via
the two-body scattering relation $A(\ldots k_j,k_i
\ldots)=\frac{k_j-k_i+\mathrm{i}c'}{k_j-k_i-\mathrm{i}c'}A(\ldots
k_i,k_j \ldots)$, which follows from  the discontinuity condition  in the
derivative of the wave function and the condition to ensure a continuous probability
density with regard to the eigenstate (\ref{state}).
Here the anyonic parameter $\kappa$ and the dynamical interaction $c$ are inextricably
related via the effective coupling constant $c'=c/\cos(\kappa/2)$ \cite{Kundu}. 
This results in a resonance-like effect in the effective coupling constant $c'$ with
respect to the statistical interaction   around $\kappa = \pi$, see Fig.~\ref{fig:c}.

Applying the periodic BC
$ \chi(x_1=0,x_2\ldots x_N)= \chi(x_2 \ldots x_N,x_1=L)$, 
leads to the eigenvalue $E=\sum_{j=1}^N k_j^2$ 
where the individual quasimomenta $k_j$ satisfy the Bethe ansatz equations (BAE)
\begin{equation}
{\mathrm e}^{\mathrm{i}k_jL}=-{\mathrm e}^{\mathrm{i}\kappa(N-1)} \prod^N_{\ell = 1} 
\frac{k_j-k_\ell+\mathrm{i}\, c'}{k_j-k_\ell-\mathrm{i}\, c'} 
\label{BA}
\end{equation}
for $j = 1,\ldots, N$.
These equations differ slightly to those of Ref.~\cite{Kundu}.  
The Bethe roots $k_j$ are real for $c'> 0$, but may become complex for
$c'< 0$. In this way we see that the $1$D
interacting anyons with periodic BC are equivalent to  a $1$D
$\delta$-function interacting Bose gas with
twisted BC, where the interaction strength is tuned via $c'$.  

\begin{center}
\begin{figure}[t]
\includegraphics[width=0.95\linewidth]{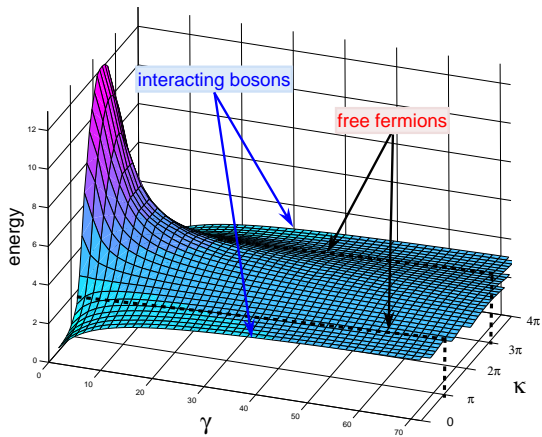}
\caption{
The lowest energy $e(\gamma,\kappa)$ in units of $n^2$ obtained from the 
integral equation (\ref{BA2}). 
For $0 < \kappa < \pi$ the energy curve interpolates between interacting bosons at 
$\kappa = 0$ and free fermions at $\kappa=\pi$ (dashed line).
For $\pi < \kappa <  3\pi$ the effective interaction $c'$ is negative.
The super TG gas-like phase is seen in the strong coupling limit $\gamma \gg 1$.
For  $3\pi < \kappa <  4\pi$ the interpolation is from free fermions to interacting bosons.
}
\label{fig:3D-E}
\end{figure}
\end{center}

For $\kappa =0$ the BAE (\ref{BA}) reduce to those of the $1$D Lieb-Liniger
Bose gas \cite{LL}. 
When $\kappa =\pi$ the BAE characterize free fermions. 
When $c=0$ the anyons may collapse into a condensation state with
purely anyonic statistical interaction.
In general the extra phase factor in the BAE (\ref{BA}),
picking up the statistical interaction during the scattering process, 
shifts the system into higher excitation states, as if there exists a
self-sustained Aharonov-Bohm-like flux \cite{note2}. 
The total  momentum is $p=N(N-1)\kappa/L+2d\pi/L$, where $d$ is  an
arbitrary integer.
In minimizing the energy we consider $\kappa(N-1) = \nu$ (mod $2\pi$) in the phase
factor with $-\pi \leq \nu \leq \pi$. 
Each  quasimomentum $k_j$ shifts to $k_j+\nu/L$ in the ground state. 
In the  thermodynamic limit, the lowest energy  is given by
  $E=N(n^2e(\gamma,\kappa)+\nu^2/L^2)$ where
  $e(\gamma,\kappa)=\frac{\gamma^3}{\lambda^3}\int_{-1}^1g(x)x^2dx$. 
The  root density $g(x)$ and the parameter $\lambda =c/Q$,
where $Q$ is the cut-off momentum, are
determined by Lieb-Liniger type integral  equations
\begin{eqnarray}
 g(x)&=&\frac{1}{2\pi}+\frac{\lambda \cos({\kappa}/{2})
  }{\pi}\int_{-1}^{1}\frac{g(y)dy}{\lambda^2+\cos^2(\frac{\kappa}{2})(x-y)^2} \nonumber\\
  \lambda&=& \gamma \int_{-1}^1g(x)dx. 
\label{BA2}
\end{eqnarray}
Fig.~\ref{fig:3D-E} shows the energy $e(\gamma,\kappa)$ 
evaluated numerically from (\ref{BA2}) for $\gamma >0$ and $\kappa \in [0,4\pi]$.

{\em Low energy behaviour.}
The low energy behaviour provides significant insight into the nature of the anyonic statistics 
interaction and the dynamical interaction. 
In this model the effective coupling constant $c'$ implements the
transmutation between statistical and dynamical interactions. 
In the weak coupling limit $\gamma \ll \cos(\kappa/2)$, the leading term for the
lowest energy per particle is obtained from the BAE (\ref{BA}) as
$E/N=(N-1) c'/L +\nu^2/L^2$. 
Here the anyonic statistics shift the energy upwards and the energy 
increases faster than the ground state energy of the pure Bose gas as $\gamma$ increases. 
The fractional statistics are mutual, i.e., the Haldane exclusion parameter discussed below is not a constant.

The experimental realization of the TG gas \cite{T-G} 
has shed further light on the quantum nature of $1$D many-body systems.
In particular, on the fermionization of bosons in $1$D, which can be experimentally realized via tuning 
the interaction strength.
The generalized exclusion statistics vary from Bose statistics to
Fermi statistics during the fermionization process.
This may provide opportunities to investigate generalized exclusion statistics in future experiments.
In the TG regime, i.e., $\gamma \gg 1$, 
the anyonic statistical interaction may trigger another regime in which 
the density-density correlations are more strongly correlated than in the
TG gas, namely the super TG gas \cite{Astr,BMGO}. 
Here this super TG gas-like phase is seen to be stable because there exists a large kinetic energy
inherited from the TG phase as the anyonic parameter $\kappa$ is
tuned smoothly from $\kappa < \pi$ to $\kappa > \pi$.  
In this way, the hard core behaviour of the particles with Fermi-like pressure prevents the collapse of
the super TG phase.
The statistics-induced super TG phase ($\pi<\kappa <3\pi$) appears only in the strong coupling limit.
It may become unstable as the interaction strength becomes weaker due to the appearance of bound states.
In general the anyonic parameter $\kappa$ implements a range of different statistical phases,  
from the Bose gas to the TG gas, from the free Fermi phase to the super TG phase.

In the TG regime the lowest energy per particle is 
$E_0/N \approx
\frac{\pi^2}{3L^2}\left(N^2-1\right)\left(1-{4 \gamma^{-1} \cos({\kappa}/{2})}\right) + {\nu^2}/{L^2}$ 
with the impenetrable fermionic distribution $\left\{\pm k_{2m}, m=1,\ldots, {(N-1)}/{2}\right\}$ 
for odd $N$, where
$k_{\ell} =\frac{\ell\pi}{L} \left(1-{2 \gamma^{-1} \cos({\kappa}/{2})} \right) + \nu/L $.  
In the thermodynamic limit and at zero temperature, the last term in $E_0/N$ can be
ignored compared to the kinetic and interaction energies. 
Now consider the effect of the anyonic statistical interaction on the linear dispersion relation for the lowest excitation. 
The elementary lowest excitation is obtained by moving the largest quasimomentum $k_N$ from 
the Dirac sea to $k_N+p$. 
To $O(p^2)$,  the low-lying excitation close to the Fermi point
\begin{equation}
E=E_0+p\left[\frac{2(N-1)\pi}{L}\left(1-\frac{4\cos({\kappa}/{2})}{\gamma}
  \right)+\frac{2\nu}{L}
  \right] 
\end{equation}
follows from the discrete BAE (\ref{BA}).
The dispersion remains linear, with sound velocity
$v_c=v_{F}(1-{4 \gamma^{-1} \cos(\kappa/2)})$ as $p\to 0$  in the thermodynamic limit. 
In the above equation, the term $2p\nu/L$  is irrelevant.  
Here the Fermi velocity $v_{\rm F}=2\pi n$. 
The finite-size corrections to the lowest energy in the
thermodynamic limit for strong coupling are directly given by 
$E_0(N,L)-Le^{(\infty)}_0=-\frac{ \pi C v_c}{6L} +O(1/L^2)$ with central charge $C=1$.

The thermodynamic BAE (TBA) is the key equation for understanding 
Haldane exclusion statistics of the model. 
Following the Yang-Yang approach \cite{Y-Y}, the TBA and the thermodynamic potential are given by
\begin{equation}
\epsilon (k)=
\epsilon^0(k) -\mu-\frac{T}{2\pi}\int_{-\infty}^{\infty}dk^{'}\theta
^{'}(k-k^{'})\ln(1+{\mathrm e}^{-\frac{\epsilon(k^{'})}{T}}) \label{TBA}
\end{equation}
\begin{equation}
\Omega
=-\frac{T}{2\pi}\int_{-\infty}^{\infty} dk \ln(1+{\mathrm e}^{-\frac{\epsilon(k)}{T}}). \label{TBA-P}
\end{equation}
Here $T$ is the temperature, $\epsilon(k)$ is the dressed energy, 
$\theta^{'}(x)=\frac{2c\cos(\kappa/2)}{c^2+\cos^2(\kappa/2)x^2}$ and
$\epsilon^0(k)=(k+\nu/L)^2$. 
In general this TBA result is only valid for the case $c'>0$. 
We also consider the TBA to be valid for the super TG gas phase, when bound states do not form. 


{\em Haldane exclusion statistics.} 
The crucial point of Haldane
exclusion statistics is that the number of available
single-particle states of species $i$, denoted by $d_i$, depends on the
number of other species $\left\{N_j\right\}$ when adding
one particle of the $i$th species to the system while keeping the
boundary conditions unchanged \cite{Haldane}. 
Following Refs.~\cite{Wu,Sen-Wilczek}, we define
\begin{equation}
d_i(\left\{N_j\right\})=G_i^0-\sum_j\alpha_{ij}N_j.\label{d}
\end{equation}
Here $G_i^0 = d_i(\left\{0 \right\})$ is the number of available single particle states 
with no particles present in the system, called the bare number of available single particle states. 
Haldane \cite{Haldane} defined the fractional statistical
interactions $\alpha_{ij}$ through a linear relation ${\Delta d_i}/{\Delta N_j}=-\alpha _{ij}$ 
with total number of particles $N=\sum_jN_j$. 
As remarked in Ref.~\cite{Wu}, this definition allows different species
to refer to identical particles with different momenta. 
The total energy is given by 
$E=\sum_iN_i \epsilon_i$, where $\epsilon_i$ is the energy of a particle of species $i$. 
For the ideal gas with no mutual statistics $\alpha_{ij}=\alpha \delta _{ij}$. 
The statistical distribution is then given by $n_i={1}/{({\mathrm e}^{\bar{\epsilon}_i/T}+\alpha})$ where
the function $\bar{\epsilon}_i$ satisfies \cite{Wu}
\begin{equation}
\bar{\epsilon}_i+T(1-\alpha)\ln(1+{\mathrm e}^{\bar{\epsilon}_i/T})=\epsilon_i-\mu.\label{state-e}
\end{equation}
The exclusion statistics are clearly seen from this relation.
For instance, $\alpha =0$ and $\alpha =1$ are Bose and Fermi statistics, respectively.

In order to apply Haldane statistics to the anyon model (\ref{Ham1}), 
we take a similar approach as that used for the interacting Bose gas \cite{Wu}. 
At zero temperature  there are no holes in the ground state. 
Excitations arise from moving quasimomenta out of the Dirac sea. 
Particle excitations leave $d_i$ holes in a momentum interval $\Delta k_i$.
Thus all accessible states in $\Delta k_i$ are $D_i
=L(\rho+\rho_{\rm h})\Delta k_i$.
$\rho$ and $\rho_{\rm h}$ are the density of occupied states and the
density of holes in interval $\Delta k_i$,
respectively. 
For an arbitrary state, the BAE (\ref{BA}) become 
\begin{equation}
\rho+\rho_{\rm h}=\frac{1}{2\pi}+\frac{1}{2\pi}\int
_{-\infty}^{\infty}dk'\theta'(k-k')\rho(k'). \label{d-BA}
\end{equation} 
On the other hand, from the definition $d_i$ in (\ref{d}), we have $G^0_i=\Delta k_iL/2\pi$,
so substituting $d_i$ into $D_i=d_i+N_i-1$ 
in the thermodynamic limit gives 
\begin{equation}
\rho+\rho_{\rm h}=\frac{1}{2\pi}+\int_{-\infty}^{\infty} \left(\delta(k,k')-\alpha(k,k')\right)\rho(k')dk'.\label{F-R}
\end{equation}
Comparison of (\ref{d-BA}) and (\ref{F-R}) thus gives \cite{note}
\begin{equation}
\alpha _{ij} := \alpha (k,k') =\delta(k,k')-\frac{1}{2\pi}\theta'(k-k'). 
\end{equation}

It is clearly seen that the leading order of the off-diagonal contribution to $\alpha (k,k')$ is proportional to
$(k-k')^2/c^3$ at low temperatures as $c\to \infty$. 
Here we require that the dynamical interaction $c$ overwhelms the thermal fluctuations. 
It follows that the exclusion statistics $\alpha(k,k') \approx  \alpha \, \delta (k,k')$
are independent of the quasimomenta at low temperatures.
{}From (\ref{F-R}) and the root distributions for the ground state, we
thus find the Haldane exclusion statistics parameter $\alpha \approx
1-2 \gamma^{-1} \cos(\kappa/2)$.  
The above relations suggest that $\alpha =-\Delta \rho_h(k)/\Delta \rho(k)$.
The meaning of 
$\alpha$ is that one particle excitation is  accompanied by
$\alpha$  hole excitations. 
It is interesting to note that the super TG phase corresponds to
exclusion statistics with $\alpha >1$ as $\pi < \kappa< 3\pi$. 
{}From the roots of the BAE (\ref{BA}) we have  the relation 
$\Delta k_i = k_{i+1}-k_i=\frac{2\pi}{L}(\alpha+\ell)$, where 
$\ell $ is a positive integer for an arbitrary state. 
This relation was also noticed in the study of exclusion statistics in 
the Calogero-Sutherland model \cite{Ha,Ha2} and provides further
evidence for its universality \cite{Sutherland}. 
Further, the quasiparticle dispersion relation can be expressed as 
$E-E_0\approx (p^2+2k_{\rm F}p)\alpha^2$ as $p\to 0$,  
where $k_{\rm F}= n \pi$ is the Fermi momentum.

It is clear to see from (\ref{state-e}) that as $T \to 0$,  
$n(k)\approx1/\alpha $ if $\epsilon^0(k)\leq \mu_0$, where  
$\mu_0 \approx k_F^2\alpha ^2$ 
is the chemical potential at $T=0$. 
Also $n(k)=0$ if $\epsilon^0(k) > \mu_0$. 
These results coincide with the BAE result for $\rho(k)$ via the relation 
$\rho (k)=g^0(k)n(k)$ with $g^0(k)=1/2\pi$ at zero temperature.
On the other hand, in the weak coupling limit $c'\to 0$, the quasimomentum
distributions are not uniform. 
The mutual statistics are governed by $\alpha (k,k')\approx -\frac{c'}{\pi(k-k')^2}$. 
Thus if $c=0$ one recovers Bose statistics.

To conclude, we have derived the low energy properties and the 
Haldane exclusion statistics of the $1$D integrable anyon gas.
We have given analytic expressions for the ground state energy, dispersion relations,
finite-size corrections and the Haldane statistical parameter and have
explicitly considered the strong and weak coupling regimes. 
We found that the anyonic statistical interaction and the dynamical
interaction implement a continuous range of Haldane exclusion
statistics, from Bose statistics to Fermi statistics and beyond.

\noindent
{\em Acknowledgements.} 
The authors thank C. Lee, H.-Q. Zhou and M. Bortz for helpful discussions and
the Australian Research Council for support.

\end{document}